\documentstyle[12pt]{article}
\begin{document}
\title{The Classical and Quantum Analysis of a Charged Particle on the 
Spacetime Produced by a Global Monopole.}
\author{A. A.Rodrigues Sobreira  and E. R. Bezerra de Mello \thanks{E-mail: 
emello@fisica.ufpb.br}\\
Departamento de F\'{\i}sica-CCEN\\
Universidade Federal da Para\'{\i}ba\\
58.059-970, J. Pessoa, PB\\
C. Postal 5.008\\
Brazil}
\maketitle
\begin {abstract}
We study the classical and quantum motion of a relativistic charged
particle on the spacetime produced by a global monopole. The self-potential
, which is present in this spacetime, is considered as an external
electrostatic potential. We obtain classical orbits and quantum states for
a  spin-1/2 and  spin-0  particles.
\end{abstract}
PACS numbers: 03.65.Ge, 03.65.Pm, 03.20.+i
\newpage
\renewcommand{\thesection}{\arabic{section}.}
\section{Introduction}

$        $

Global monopoles are heavy objects probably formed in the early Universe by
the phase transition that occcur in a system composed of a self-coupling 
scalar field triplet $ \phi^{a} $ whose its original global symmetry 
$ 0(3) $ is spontaneously broken to $ U(1) $.

The simplest model that gives rise to a global monopole is described by
the Lagrangian density below
\begin{eqnarray} 
\label {01}
{\cal{L} } = \frac{1}{2}({\partial_ \mu}\Phi^a )({\partial^\mu}\Phi^a)- 
              \frac{\lambda}{4}(\Phi^a\Phi^a-\eta^2)^2
\end{eqnarray}

Coupling\,\,\,\, this\,\, matter\,\, field with the\,\,\, Einstein\,\, 
equation,\,\,\, Barriola and Vilenkin $\cite{barriola}$ have shown that the 
effect produced by this object in the geometry can be approximately 
represented by a solid angle deficit in $ ( 3+1 )$-dimensional spacetime, 
whose the line element of this manifold is given by
\begin{eqnarray}
\label{02}
ds^{2} = -dt^{2} + \frac{dr^{2}}{\alpha^{2}} + r^{2}(d\theta^{2} + 
sen^{2}\theta d\varphi^{2}),
\end{eqnarray}
where the parameter $ \alpha^{2} = 1 - 8{\pi}G{\eta}^{2} $  is smaller  
than one and depends on the energy scale  where the global symmetry is 
spontaneously broken. The area of a sphere of unit radius in this manifold 
is $ 4\pi\alpha^{2}. $

Recently has been found that a charged particle placed in the spacetime
produced by a global monopole becomes subjected to a repulsive
electrostatic self-potential$ \cite{eugenio}$ given by
\begin{eqnarray}
\label{3}
U = \frac{K}{r}
\end{eqnarray}
where r is the radial distance from the particle to the monopole and
\begin{eqnarray}
\label{4}
K = \frac{{q^2}{S(\alpha)}}{2} > 0 
\end{eqnarray}
being  $ S(\alpha) $ a numerical factor which is positive for $ \alpha < 1
$ and negative for $ \alpha > 1 $. This self-potential is consequence of
the topology of this spacetime and also its non-vanishing curvature.

Here in this paper we decided to analyse the classical and quantum motion
of a charged particle in the spacetime produced by a global monopole,
taking into account the self-potential as an external electrostatic
potential in the equations of motion. So, with the objective to develop
this analysis in a more general point of view we present this paper as
follows: In  section 2, we study the classical orbits by using the 
Hamilton-Jacobi formalism. We analyse the possibility of the test particle 
become bound to the monopole considering the  unphysical case where 
$ \alpha > 1 $. In section 3, we study the quantum motion considering the 
test particle as a bosonic and fermionic one, by using, respectively, the 
Klein-Gordon and Dirac equations. Finally in section 4 we summarize our
main results and make some remarks.

\renewcommand{\thesection}{\arabic{section}.}
\section{\bf Classical Motion}

$                   $

In this section we study the relativistic classical motion of a particle
with mass $ M $ and electric charge $ q $ subjected to the electrostatic
self-potential $ (\ref{3}) $ and $ (\ref{4}) $ in the manifold described
by in $ (\ref{02}) $. In order to do that, we shall use the Hamilton-Jacobi
formalism. The first set of equations to be used is given by
\begin{eqnarray}
\label{5}
M\frac{dx^{\mu}}{d{\lambda}}+qA^{\mu}=g^{\mu\nu}\frac{{\partial}S}{{
         \partial}x^{\nu}},
\end{eqnarray}
where 
\begin{eqnarray}
\label{6}
A^{\mu}=\frac{K}{qr}\delta^{\mu}_{0},
\end{eqnarray}
being $ \lambda $ a parameter along the trajectory of the particle.

The second set is given by
\begin{eqnarray}
\label{7}
g^{{\mu}{\nu}}\left(\frac{{\partial}S}{{\partial}{x^{\mu}}}-q{A_ {\mu}}
        \right)\left(\frac{{\partial}S}{{\partial}{x^{\nu}}}-q{A_{\nu}}
        \right)=-M^{2}.
\end{eqnarray}

For this system we shall employ the spherically symmetric ansatz
for the functional $ S $.
\begin{eqnarray}
\label{8}
S=-Et + R(r) + \Theta(\theta) + L_{Z}\varphi,
\end{eqnarray}
where $ R(r) $ and $\Theta(\theta) $ are unknown functions and $E$ and 
$ L_{Z} $  constants of motion.

Substituting $ (\ref{8})$ into $ (\ref{7}) $ we get the following set of
differential equations:
\begin{eqnarray}
\label{9}
\left(\frac{d{\Theta}}{d{\theta}}\right)^{2} = L^{2} - 
   \frac{L_{z}^{2}}{sin^{2}{\theta}},
\end{eqnarray}
and
\begin{eqnarray}
\label{10}
\alpha^{2}\left(\frac{d R}{d r}\right)^{2} = \left(-E + 
   \frac{K}{r}\right)^{2} - M^{2} - \frac{L^{2}}{r^{2}},
\end{eqnarray}
where $ L^{2} $ is positive constant.
\\     \\

We also can write
\begin{eqnarray}
\label{11}
M \frac{d r}{d {\lambda}} = \alpha^{2}\left(\frac{d R}{d r}\right),
\end{eqnarray}
which, combining with $ (\ref{10}) $ results in
\hspace{1cm}
\begin{eqnarray}
\label{12}
\frac{M^{2}}{\alpha^{2}}\left(\frac{d r}{d {\lambda}}\right)^{2} = 
   \left(E - \frac{K}{r}\right)^{2} - M^{2} - \frac{L^{2}}{r^{2}}.
\end{eqnarray}
Also, from $ (\ref{5}) $ we get
\begin{eqnarray}
\label{13}
M\left(\frac{d\theta}{d\lambda}\right) = \frac{1}{r^{2}}\left(\frac{d
        \Theta}{d\theta}\right)
\end{eqnarray}
and
\begin{eqnarray}
\label{14}
M\left(\frac{d\varphi}{d\lambda}\right) = \frac{1}{r^{2}sin^{2}\theta}
        {L_{Z}}.
\end{eqnarray}

Combining $ (\ref{13}) $ and $ (\ref{14}) $ we obtain a differential
equation relating the two angular variables, which by integration yields
\\      \\
\begin{eqnarray}
\label{15}
cot^{2}\theta = (\chi^{2} - 1 )sin^{2}\varphi,
\end{eqnarray}
\\       \\
with $ \chi=\frac{L}{L_{Z}}$.We can see from the equation above that the 
relation between both angular variables does not depend on the parameter
$ \alpha $.

In order to analyze the dependence of the radial coordinate on the angular 
ones, what we call by equation of the classical orbits, we shall derive the 
function $ r(\varphi) $.\, In order to simplify our calculation only,
let us consider  the surface $ \theta = \pi/2, $ and defining
$ u = 1/r $ we get after some steps:
\begin{eqnarray}
\label{16}
\frac{1}{\alpha^{2}\chi^{2}}\left(\frac{du}{d\varphi}\right)^{2} 
+ u^{2} = \frac{1}{L^{2}}\left[(E - Ku)^{2} - M^{2}\right],
\end{eqnarray}
whose solutions are
\begin{eqnarray}
\label{17}
u(\varphi) = \frac{E K}{L^{2} - K^{2}}\!\!\! +\! \frac{\sqrt{(E^{2} - 
  M^{2})L^{2} + M^{2}K^{2}}}{L^{2}- K^{2}}{cos}\left[\frac{\sqrt{L^{2} 
    - K^{2}}}{L_{Z}}{\alpha}(\varphi - \varphi_{0})\right],
\end{eqnarray}
for $ L^{2} > K^{2}, $
\begin{eqnarray}
\label{18}
u(\varphi)& =& \frac{E K}{K^{2} - L^{2}} {\pm } \frac{\sqrt{(E^{2} - 
   M^{2})L^{2} + M^{2}K^{2}}}{K^{2} - L^{2}}\times\nonumber\\
       && \cosh\left[\frac{\sqrt{K^{2} - L^{2}}}{L_{Z}}{\alpha}
                   (\varphi - \varphi_{0})\right],
\end{eqnarray}
for $ L^{2} < K^{2}, $ and
\begin{eqnarray}
\label{19}
u(\varphi) = \frac{M^{2}}{2E K} - \frac{E \alpha^{2} \chi^{2}(\varphi 
           - \varphi_{0})^{2}}{2K}
\end{eqnarray}
for $ L^{2} = K^{2}.$

From the equations above we can see that the trajectories described by $ 
(\ref{17}),\,\,\, $ for the attractive case,\,\,\, $ K < 0 $ \,\,\,
( which happens for the unphysical situation when $\alpha > 1 $),
will be bounded from below by some positive value if $  E < M ,$
so $r$ will be bounded from above.\, If $ E > M,$ $ r $  can be infinity.
The trajectories given by $ (\ref{18}) $ and $ (\ref{19}) $ are spiral
ones, and  $r$ goes to zero when $ \varphi $ goes to infinity. In
relativistic classical mechanics finite trajectories in general are not
closed but rather rosetteshaped $ \cite{landau}$. The former case presents
period given by
\begin{eqnarray}
\label{20}
T = \frac{2\pi L_{Z}}{\alpha{\sqrt{L^{2} - K^{2}}}}.
\end{eqnarray}
which clearly on the parameter $ \alpha $.

There is also a constant solution obtained by setting $ du/d\varphi = 0 $ 
in $ (\ref{16}) $.\,This circular motion ocurs in relativistic theory just 
as in classical one.

The relation between $ u = 1/r $ and\,\, the polar angle\,\,\, $ \theta $ 
can also be obtained. After some minor development from the previous 
equations \,\,we get for\,\, $ L^{2} > K^{2} $,
\begin{eqnarray}
\label{21}
u(\psi)&=& -\frac{E K}{L^{2} - K^{2}} + \frac{\sqrt{(E^{2} - M^{2})L^{2} 
                   + M^{2}K^{2}}}{L^{2}- K^{2}}\times\nonumber\\
                     &&{cos}\left[\frac{\sqrt{L^{2} 
                     - K^{2}}}{L_{Z}}{\alpha}(\psi - \psi_{0})\right],
\end{eqnarray}
where the new angular variable $ \psi $ is expressed in terms of $ \theta $
by
\begin{eqnarray}
\label{22}
\psi = - \frac{1}{\chi}{arcsin}\left(\frac{\chi{cos}\theta}{\sqrt{\chi^{2} 
             - 1 }}\right).
\end{eqnarray}

We also should note that the equations above depend on the parameter
$ \alpha $, however these two last relations differ from the previous ones
envolving the azimuthal angle, where it appears as a factor multiplyng
$ \varphi $.

Combining equation $ (\ref{11}) $ with the $ \mu = 0 $ component of $ 
(\ref{5}) $ we can obtain an expression for the radial velocity for the 
test particle.
\begin{eqnarray}
\label{23}
\frac{d r }{d t} = \frac{\alpha^{2}}{E - K/r}\frac{d R}{d r}.
\end{eqnarray}

The turning points of the trajectory are given by setting $ dr/dt = 0 $.\,
As  consequence we get the potential curves given by
\begin{eqnarray}
\label{24}
\frac{E}{M} = \left(1 + \frac{L^{2}}{M^{2}r^{2}}\right)^{1/2}  + \frac{K}
             {Mr}.
\end{eqnarray}

For $ K > 0 , $ which means the physical situation with $ \alpha < 1 $
the right hand side of the above expression presents no extremals, so the
particle cannot be bound by the monopole; however for $ K < 0 $, which is
associeted with the unphysical situation with $ \alpha > 1, $  it is
possible to have $ E/M < 1 $ for a finite radius in $ (\ref{24})$.\,
(In this case the  potential curves above coincides with the circular 
trajectory mentioned previously). The situation where the self-potential 
is absent, $K = 0 $ , has been analysed in $ \cite{chakraborty}$.\,There 
it was found no possibility of the test particle be trapped by the
monopole.

After this analysis about the classical motion of the test particle in this
manifold, where exact solutions were obtained, we shall leave for the next
section the relativistic quantum motion study for this system.
\renewcommand{\thesection}{\arabic{section}.}
\section{\bf Quantum Motion}
$        $

This section is devoted to the relativistic quantum analysis of scattering
and bound states of a massive charged test particle in the spacetime metric
given by $ (\ref{02}), $  considering the self-potential as an external
electrostatic potential. In this sense this section is a natural extension 
of the previous one, taking into account the quantum effects. In order to 
make this analysis complete, we shall consider the cases where the test 
particle is a bosonic and fermionic particle. We start first this analysis 
considering the former case.

$         $\\

A) {\bf Bosonic Case}
\\         \\

The Klein-Gordon equation written in a covariant form in the presence of an
external four-vector potential $ A^{\mu} $ reads $ \cite{gross} $
\begin{eqnarray}
\label{25}
[\Box- \frac{i q A^{\mu}}{\sqrt{-g}}(\partial_{\mu} \sqrt{-g}) 
          -iq(\partial_{\mu} A^{\mu})-2iq A^{\mu}{\partial}_{\mu} 
              - q^{2}A^{\mu}A_{\mu} - M^{2}]\varphi(x) = 0,
\end{eqnarray}
with
\begin{eqnarray}
\label{26}
\Box \varphi(x) = \frac{1}{\sqrt{-g}}\partial_{\mu}[\sqrt{-g}g^{\mu\nu}
                    \partial_{\nu}\varphi(x)],
\end{eqnarray}
being $ g = det(g_{\mu\nu}). $

We can also consider in our analysis the so called non-minimal coupling
between the scalar field $ \varphi(x) $ with the geometry of the manifold
itself. This term, which is invariant, is expressed by $ \Upsilon $ R $
\varphi, $ where the coeficient  $\Upsilon $ is an arbitrary coupling 
constant and R the scalar curvature.\,(For the massless case the conformal
coupling constant is $ \Upsilon = 1/6 $.)

The inclusion of the non-minimal coupling in our formalism, does not make 
our analysis more complicate.The reason for this fact is that the scalar 
curvature assaciated with the space-time produced by a global monopole is
$ R = 2(1 - \alpha^{2})/r^{2}, $ which presents the similar behaviour of 
the centrifugal potential energy.

Now, writing down the Klein-Gordon equation in the manifold described by 
$ (\ref{02}), $ including the  self-potential for a charged particle given 
by $ (\ref{6}) $ and also the non-minimal coupling, we get from $ (\ref
{25})$:
\begin{eqnarray}
\label{27}
[-\partial^{2}_{t} + \frac{\alpha^{2}}{r^{2}}{\partial_{r}}(r^{2}{\partial
          _{r}}) - \frac{\bf L^{2}}{r^{2}} - 2i{\frac{K}{r}}{\partial_{t}} 
             + \frac{K^{2}}{r^{2}}-M^{2} - {\Upsilon}\frac{\eta}{r^{2}}]
                   \varphi(x) = 0,
\end{eqnarray}
where
\begin{eqnarray}
\label{28}
\eta = 2(1 - \alpha^{2})
\end{eqnarray}
and $ {\bf L} $ the usual orbital angular momentum operator.

Because our metric tensor is a static one and the self-potential is time 
independent, we shall adopt for the wave function the form  below
\\        \\
\begin{eqnarray}
\label{29}
\varphi(x) = e^{-i E t}R(r) Y^{m}_{l}(\theta, \varphi),
\end{eqnarray}
\\       \\
where E is the energy of the particle. Substituting  $ (\ref{29}) $ 
into $ (\ref{27}) $
we get the radial Klein-Gordon equation.
\begin{eqnarray}
\label{30}
\frac{\alpha^{2}}{r^{2}}\frac{d}{d r} \left(r^{2}\frac{d R}{d r}\right) 
           - [l(l + 1) + \Upsilon\eta - K^{2}] \frac{R}{r^{2}} - \frac{2EK}
           {r}R + (E^{2} - M^{2})R = 0 ,
\end{eqnarray}
where we can identify the equation above with the Schrodinger one,
associating,
\begin{eqnarray}
{\bf L}^{2} \rightarrow \frac{{\bf L^{2}} + \Upsilon\eta - K^{2}}{\alpha^
{2}},
\nonumber
\end{eqnarray}
\begin{eqnarray}
V(r) \rightarrow \frac{EK}{M\alpha^{2}r},
\nonumber
\end{eqnarray}
and
\begin{eqnarray}
E \rightarrow \frac{E^{2}- M^{2}}{2M\alpha^{2}}.
\nonumber
\end{eqnarray}

We shall analyse in this section scattering states mainly; however if we 
admit the possibility of the parameter $ \alpha^{2} $ in $ (\ref{02}) $
be bigger than one, the self-potential will be an attractive one, so bound
states can be present in this case. Analysing the behaviour of the wave
functions near the origin, and assuming $ E^{2} > M^{2}, $ well defined 
scattering states can be obtained expressed in terms of confluent 
hypergeometric functions as :
\begin{eqnarray}
\label{31}
R(r) = C r^{\lambda_{l}} e^{i {\kappa} r} {_{1}F_{1}}(\lambda_{l} + 1 
              +  i \beta ,2(\lambda_{l} + 1 ); -2 i {\kappa} r),
\end{eqnarray}
where $ \lambda_{l} = -\frac{1}{2} + \frac{\sqrt{\alpha^{2} + 4[l(l+1) + 
          \Upsilon\eta - K^{2}]}}{2\alpha},\,\,\,\, \, \,\kappa = \frac{
          \sqrt{E^{2} -M^{2}}}{\alpha}\,$  and $ \, \beta = \frac{E K}
          {\alpha^{2}\kappa}$.
\\       \\

Following the standard procedure $ \cite{schiff} $ we can take the 
asymptotic form of the confluent hypergeometric function  obtain the
long distance behaviour for the radial function.
\\       \\
\begin{eqnarray}
\label{32}
R(r) \sim C \frac{{\Gamma(2\lambda_{l} + 2 )}e^{\beta{\pi/2}}}
              {\mid\Gamma(\lambda_{l} +1 + i\beta)\mid }\frac{{cos}
                    (\kappa r - {(\pi}/2)(\lambda_{l} + 1) 
                        - {\beta}{ln}(2 \kappa r)+\gamma_{l})}{\kappa r},
\end{eqnarray}
\\       \\
where $ \gamma_{l} = {arg}\Gamma(\lambda_{l} + 1 + i{\beta}). $

From the above equation it is possible to obtain the phase shift $ \delta_
{l}, $ which is the most relevant parameter in the calculation of the
scattering amplitude $ \cite{schiff}.$
\begin{eqnarray}
\label{33}
\delta_{l} = \frac{\pi}{2}(l - \lambda_{l}) + \gamma_{l}.
\end{eqnarray}

As we can see from the result above, the phase shift presents two 
contributions :\, $ (i)$ From the modification of the effective angular 
quantum number    $ \lambda_{l} $ due to the geometry of the manifold
itself and $ (ii) $ from the presence of the repulsive self-interaction
term as another indirect consequence of this non-trivial topology. Taking
$ K = \eta = 0 $ in $ (\ref{33}) $ we reobtain the result found in the
$ \cite{mazur}.$

Now, if we are inclined to consider the possibility of the test particle
to be bound by the global monopole, we have to assume that the parameter
$ \alpha^{2} $ in $ (\ref{02}) $ is bigger than one. In this case the
induced self-interaction is attractive and bound states can be obtained
by taking in $ (\ref{30}) ,  E^{2} < M^{2} $ and $ K = - \mid K \mid. $ 
Again imposing appropriate boundary condition on the solutions, 
they can be expressed as follows :
\\      \\
\begin{eqnarray}
\label{34}
R(r) = C {\lambda_{l}} e^{-{\kappa} r} {_{1}F_{1}}(\lambda _{l}+ 1 - \xi, 
             2(\lambda_{l} + 1 ); 2 {\kappa} r)
\end{eqnarray}
where $ \kappa = \frac{\sqrt{M^{2} - E^{2}}}{\alpha} $ and $ \xi = 
\frac{E |K|}{\alpha^{2}\kappa}$.

In order to obtain bound states we have to choose appropriately the
parameters to terminate the series in $ (\ref{34}) $.  Admitting a
polynomial of degree $ n $  for the hypergeometric function, we must impose
the condition
\begin{eqnarray}
\label{35}
\frac{E {\mid }K {\mid}}{\alpha^{2} \kappa} - \lambda_{ l} - 1 = n.
\end{eqnarray}

With this condition we get discret values for the sel-energy given by :
\begin{eqnarray}
\label{36}
E^{(\alpha)}_{n,l} = M \left[ \frac{\alpha^{2}(n + \lambda_{l}+ 1)^{2}}
           {K^{2} + \alpha^{2}(n + \lambda_{l}+1)^{2}}\right]^{1/2},
\end{eqnarray}
with $ n = 0,1,2... $

Observing the expression above we  call attention for two interesting
results:\, $(i) $ The first one refers to the degeneracy problem. Because 
$ \alpha \not= 1, $ the self-energy depends on $ n + \lambda_{l} $ which 
is not in general an integer number, this reduce the degree of the
degeneracy of our solutions.\, $ (ii) $ The second one refers with self-
energy itself. As we can $  E^{(\alpha)}_{n,l} $ depends on the parameter
$ K^{2} $ which in general can be a very small quantity, so expanding  $
(\ref{36}) $ in powers of $ K^{2} $ we get the non-relativistic limit of
the energy  \footnote{This result would be obtained if we have used the
Schrodinger equation defined in the geometry defined in  (\ref{02}), in
the presence of an external electrostatic potential (\ref{3}). (See Ref. $
[2]$)}
\begin{eqnarray}
\label{37}
E_{n,l} \simeq M \left[1 - \frac{K^{2}}{2\alpha^{2}(n+\lambda_{l}+1)^{2}}
     \right].
\end{eqnarray}
$       $\\

B)  {\bf Fermionic Case}
\\       \\

The Dirac equation written in a covariant form and in presence of an
external four-vector potential reads
\begin{eqnarray}
\label{38}
[ i\gamma^{\mu}(x)(\partial_{\mu} + i q A_{\mu} - \Gamma_{\mu}(x)) 
           - M]\Psi(x) = 0, 
\end{eqnarray}
where $ \Gamma_{\mu}(x) $ is the spinorial connection coeficient given in
terms of the tetrads and Christoffel symbols
\begin{equation}
\label{39}
\Gamma_{\mu} = - \frac{1}{4} \gamma^{(a)} \gamma^{(b)} e^{\nu}_{(a)}
              (\partial_{\mu} e_{(b) \nu} - \Gamma^{\lambda}_{\nu\mu}e_{(b)
               \lambda})
\end{equation}

The generalized Dirac  matrices $ \gamma^{\mu}(x) $ are given in terms of 
the standard flat  spacetime gammas $ \gamma^{(a)} $ by the relation
\begin{equation}
\label{40}
\gamma^{\mu}(x) = e^{\mu}_{(a)} \gamma^{(a)} 
\end{equation}

For the metric corresponding to the global monopole we shall use the 
following tetrads :
\begin{eqnarray}
\label{41}
e^{\mu}_{(a)} = \left(
\begin{array}{cccc}
1   & 0  &  0  &  0 \\
\\       \\
0   & {\alpha}{sin}\theta{cos}\varphi & r^{-1}{cos}\theta{cos}\varphi & 
        -(r{sin}\theta)^{-1}{sin}\varphi \\
\\      \\
0   &  {\alpha}{sin}\theta{sin}\varphi & r^{-1}{cos}\theta{sin}\varphi & 
         (r{sin}\theta)^{-1}{cos}\varphi \\
\\       \\
0   &  \alpha{cos}\theta     &   -r^{-1}{sin}\theta      &    0
\end{array}
\right)
\end{eqnarray}
which obey the relation
\begin{eqnarray}
\label{42}
e^{\mu}_{(a)}e^{\nu}_{(b)}\eta^{(a)(b)} = g^{{\mu}{\nu}},
\end{eqnarray}
being  $ g^{{\mu}{\nu}}(x) $ given by $ (\ref{02}) $. (Although there is a
simpler basis tetrad to describe the metric tensor associeted with a global 
monopole, the above choice is convenient, as we shall see, in the sense
that taking $ \alpha = 1 $ we reobtain the standard expression for the
Dirac equation in a flat spacetime.) Now writting the Dirac equation in
this spacetime and in presence of the  self-potential we get:
\begin{eqnarray}
\label{43}
[i\gamma^{(0)}\partial_{t} + i{\alpha}{\gamma^{(r)}}\partial_{r} + \frac{i}
       {r}{\gamma^{(\theta)}}\partial_{\theta} + \frac{i}{r{sin}\theta}
       \gamma^{(\varphi)}\partial_{\varphi} + i\frac{(\alpha - 1)}{r}{
       \gamma^{(r)}} -\nonumber\\- q {\gamma^{(0)}}A_{0} - M ]\Psi(x) = 0,
\end{eqnarray}
where $ \gamma^{(r)} = {\vec \gamma}.\hat{r}, \gamma^{(\theta)} = {\vec
\gamma}.\hat{\theta} $ and $ \gamma^{(\varphi)} = {\vec \gamma}.\hat{
\varphi}, $ being $ \hat{r}, \hat{\theta} $ and $ \hat{\varphi} $ th
standard unit vector along the three spatial directions in the spherical
coordinates. We shall use the following representation of the flat space
$ \gamma$-matrices.
\begin{eqnarray}
\gamma^{(0)} = \left(
\begin{array}{cc}
1 & 0 \\
0 & -1
\end{array}
\right),
\nonumber
\end{eqnarray} 
\begin{eqnarray}
\gamma^{(i)} = \left(
\begin{array}{cc}
0 & \sigma^{i} \\
-\sigma^{i} & 0
\end{array}
\right)
\nonumber
\end{eqnarray} 
and for the complete set solutions to the Dirac equation
\begin{eqnarray}
\label{44}
\Psi({\bf r},t) =\frac{1}{r} \left(
\begin{array}{c}
{i}F_{{j}{m}}(r){\Phi_{{j}{m}}}({\theta},{\varphi})\\
\\        \\
G_{{j}{m}}(r)(\vec{\sigma}.\hat{r})\Phi_{{j}{m}}({\theta}{\varphi})
\end{array}
\right)e^{{-i}{E}{t}}
\end{eqnarray}
which presents a well defined parity under the tranformation
$ \vec{r} \rightarrow \vec{r}\,^ {'}=-\vec{r}.$
The $ \Phi_{jm}(\theta,\varphi) $ are the spinor spherical harmonics.

Substituting $ (\ref{44}) $ into $ (\ref{43}) $ and after some minor steps
we find the set of radial diferential equations,
\begin{eqnarray}
\label{45}
(E - M - K/r)F_{jm} = -\alpha \frac{dG_{jm}}{dr} + {\eta}\frac{G_{jm}}{r},
\end{eqnarray}
\begin{eqnarray}
\label{46}
(E + M - K/r)G_{jm} = \alpha \frac{dF_{jm}}{dr} + {\eta}\frac{F_{jm}}{r}
\end{eqnarray}
where $ \eta = \mp (j + 1/2)  \cite{bjorken}$. Let us first consider the 
possibility of this system, charged particle and global monopole, presents
bound states. This fact, as we have mentional previously, is only possible 
when $ K<0$, which implies $ \alpha > 1 $. (Although this situations is 
realy an unphysical one, we decided to condider it below in order to make 
our analysis a complete one). We shall try solutions of the equations 
$ (\ref{45})$ and $ (\ref{46}) $ in the form
\begin{eqnarray}
\label{47}
F_{jm}(r) = C(1 + E/M)^{1/2} e^{-{\kappa}r}({\kappa}r)^{\lambda_{j}}(F_{1}
           (r)+ F_{2}(r)),
\end{eqnarray}
and
\begin{eqnarray}
\label{48}
G_{jm}(r) = C(1 - E/M)^{1/2} e^{-{\kappa}r}({\kappa}r)^{\lambda_{j}}
            (F_{1}(r) - F_{2}(r)),
\end{eqnarray}
where $ \kappa =\frac{\sqrt{M^{2} - E^{2}}}{\alpha} $ and $ \lambda{j} =
\frac{\sqrt{(j + 1/2)^{2} - K^{2}}}{\alpha} $ .
\\      \\

Using in $ (\ref{45}) $ and $ (\ref{46}) $ $ K = -{\mid}K{\mid} $ and 
developing  intermediate calculations, we found that the solutions for the 
unknown functions $ F_{1}(r) $ and $ F_{2}(r) $ are expressed in terms of 
confluent hippergeometric functions as
\begin{eqnarray}
\label{49}
F_{2}(r) = C{_{1}F_{1}}(\lambda_{j} - \xi, 2{\lambda_{j}} + 1 ; 2{\kappa}r)
\end{eqnarray}
and 
\begin{eqnarray}
\label{50}
F_{1}(r) = C\frac{\alpha^{2}{\kappa}{\lambda_{j}} - E |K|}{M|K| 
         - {\eta}{\alpha}\kappa}{_{1}F_{1}}(\lambda_{j} - \xi + 1, 
            2{\lambda_{j}} + 1 ; 2{\kappa}r)
\end{eqnarray}
where $ \xi = E K/\alpha^ {2} \kappa .$ The expression for $ F_{1}(r) $ 
has been found using the general relations between the confluent 
hippergeometric functions $ \cite{abramowitz} $
\begin{eqnarray}
(z\frac{d}{dz} + a){_{1}F_{1}}(a,b;z) = a{_{1}F_{1}}(a +1,b;z).
\nonumber
\end{eqnarray}
So the complete expressions to the functions $ F$ and $ G $ are given by :
\begin{eqnarray}
\label{51}
F_{jm}(r) &&= C(1 + E/M)^{\frac{1}{2}}e^{-
          {\kappa}r}({\kappa}r)^{\lambda_{j}}\Bigg[{_{1}F_{1}}(\lambda_{j}
          -\xi ,2{\lambda_{j}} + 1 ; 2{\kappa}r) + \nonumber\\
             &&\left. +\frac{\alpha^{2}{\kappa}{\lambda_{j}} - E {\mid}K
             {\mid}}{M{\mid}K{\mid} - {\eta}{\alpha}\kappa}{_{1}F_{1}}(
             \lambda_{j} - \xi+ 1, 2{\lambda_{j}} + 1 ; 2{\kappa}r)\right]
\end{eqnarray}
\\       \\
and
\\       \\
\begin{eqnarray}
\label{52}
G_{jm}(r)=&&-C(1 - E/M)^{1/2} e^{-{\kappa}r}({\kappa}r)^{\lambda_{j}}
           \Bigg\{{_{1}F_{1}}(\lambda_{j} - \xi, 2{\lambda_{j}} + 1 ; 2{
           \kappa}r) -\nonumber\\ 
            && - \frac{\alpha^{2}{\kappa}{\lambda
           _{j}} - E |K|}{M|K|  -{\eta}{\alpha}{\kappa}}{_{1}F_{1}}(\lambda
           _{j} - \xi+ 1, 2{\lambda_{j}} + 1 ; 2{\kappa}r)\Bigg\}.
\end{eqnarray}

From the expressions above we can see the dependence of the radia
functions on the parameter $ \alpha $ which define this geometry and also
on the external electrostatic potential through the constant $ K$.

In order to obtain discret values for the self-energies it is necessary to
impose the vanishment  condition on the wave function when $ {\kappa}r{
\rightarrow}\infty $. From the asymptotic behaviour the confluent
hippergeometric function $ \cite{abramowitz} $ we get.
\begin{eqnarray}
\label{53}
\frac{E|K|}{\kappa{\alpha^{2}}} - \lambda_{j} =n, \;\;\;\;\;\;\;\; 
                     n=0,1,2...
\end{eqnarray}

Using the values for $ \kappa $ and $ \lambda_{j} $ given previously we 
obtain the explicit expression for the self-energy.
\begin{eqnarray}
\label{54}
E_{n,j} = {M}\left[1 + \frac{K^{2}}{(n{\alpha} + \sqrt{(j+1/2)^{2} 
               - K^{2}})^{2}}\right]^{-1/2}.
\end{eqnarray}

Also from the expression above we see the energy depend on the parameter 
$ \alpha $ in two different ways, through this parameter itself and the 
constant $ K $ indirectly. (Considering $\alpha = 1 $ and keeping the
constant $ K\not=0, $ the self-energy above reproduces the values of the
self-energies of an electron in the hidrogen atom.) The explicit normalized
eigenfunctions $ (\ref{44}) $ can also be obtained.

Let us now study scattering states. These states are obtained considering 
$ E^{2} > M^{2} $ in $ (\ref{45}) $ and $ (\ref{46}) $. This is the real
case because $ K>0. $

Again we shall try solutions  in the form :
\begin{eqnarray}
\label{55}
F_{j,m}(r) =C (1 + E/M)^{1/2} e^{i{\kappa}r}({\kappa}r)^{\lambda_{j}}
(F_{1}(r) + F_{2}(r)),
\end{eqnarray}
\begin{eqnarray}
\label{56}
G_{j,m}(r) = -C(1 - E/M)^{1/2} e^{i{\kappa}r}({\kappa}r)^{\lambda_{j}}
(F_{1}(r) - F_{2}(r)),
\end{eqnarray}
where $ \kappa =\frac{\sqrt{E^{2} - M^{2}}}{\alpha}.$

Substituting the expressions above in $ (\ref{45}) $ and $ (\ref{46}) $ and
developing some intermediate calculations we find that $ F_{1}(r) $ and
$ F_{2}(r) $ can also be expressed in terms of complex confluent
hippergeometric functions as :
\begin{eqnarray}
\label{57}
F_{2}(r) = C_{1}F_{1}(\lambda_{j} + i \beta, 2\lambda_{j} + 1 ; -2i\kappa
r)
\end{eqnarray}
and
\begin{eqnarray}
\label{58}
F_{1}(r) = C \frac{\alpha^{2}(i \beta + \lambda_{j})}{i(\frac{-MK}{\kappa} 
             + i \eta \alpha)} {_{1}F_{1}}(\lambda_{j} + 1 + i \beta, 
                 2\lambda_{j} + 1 ; -2i\kappa r),
\end{eqnarray}
where $ \beta = EK/\alpha^{2} \kappa.$

The complete expressions for the scattering wave function can be given by
\begin{eqnarray}
\label{59}
F_{j,m} =&&  C (1 + E/M )^{1/2} e^{i \kappa r}(\kappa r)^
          {\lambda_{j}}\Bigg\{{_{1}F_{1}}(\lambda_{j} + i \beta, 2\lambda_
          {j}+ 1 ; -2i\kappa r) + \nonumber\\ && +  \frac{\alpha^{2}(i
          \beta+ \lambda_{j})}{i(\frac{-MK}{\kappa} + i \eta \alpha)} {_{1}
          F_{1}}(\lambda_{j} + 1 + i \beta, 2\lambda_{j} + 1 ; -2i\kappa r)
             \Bigg\},
\end{eqnarray}
and
\begin{eqnarray}
\label{60}
G_{j,m} =&& C (1 - E/M)^{1/2}e^{i \kappa r}(\kappa r)^{\lambda_{j}}
           \Bigg\{{_{1}F_{1}}(\lambda_{j} + i \beta, 2\lambda_{j} + 1 ; 
            -2i\kappa r) - \nonumber\\ && -  \frac{\alpha^{2}(i \beta +
            \lambda_{j})}{i(\frac{-MK}{\kappa} + i \eta \alpha)} {_{1}F_
            {1}}(\lambda_{j} + 1 + i \beta, 2\lambda_{j} + 1 ; -2i\kappa r)
            \Bigg\},
\end{eqnarray}

As in the bosonic case we can also obtain the phase shift by the asymptotic 
form  of the wave function $ ( \ref{44}),$ which by its turn depend on the 
above expressions. So, let us write down the long distance behavior for $
F $ and $ G.$
\begin{eqnarray}
\label{61}
F_{jm} \approx && C \left( 1 + E/M \right) ^{\frac 12}
      \frac{\Gamma (2 \lambda _{j} +1)}{|\Gamma (\lambda _{j} + i\beta)|}
          \frac{2}{\left( -\frac{\eta}{\alpha} -i\frac{MK}{\kappa \alpha ^
          2}\right)}e^{\beta \pi /2} e^{\frac 12 (\sigma _{1} -\sigma _{2}
          )}\times \nonumber\\&&\Bigg\{\cos \left[\kappa r - \lambda _{j}
          \frac{\pi}{2} - \beta ln(2\kappa r) + \gamma _{j} +\frac{\sigma _
          {1} - \sigma _{2}}{2}\right]+\nonumber\\ && \frac{|\lambda _{j}
          + i \beta|}{2 \kappa r} \sin\left[\kappa r - \lambda _{j} \frac{
          \pi}{2} - \beta ln(2\kappa r)+ \gamma _{j} +\frac{\sigma _{1} +
          \sigma _{2}}{2}\right]+...\Bigg\}
\end{eqnarray}
and
\begin{eqnarray}
\label{62}
G_{jm} \approx && -C ( 1 - E/M) ^{\frac 12}\frac{\Gamma (2 \lambda _{j} +1)
           }{|\Gamma (\lambda _{j} + i\beta)|}\frac{2}{\left( -\frac{\eta}
           {\alpha} -i\frac{MK}{\kappa \alpha ^2}\right)}e^{\beta \pi /2}
           e^{\frac 12 (\sigma _{1} -\sigma _{2})}\times \nonumber\\&&\Bigg
           \{\sin \left[\kappa r - \lambda _{j} \frac{\pi}{2} - \beta ln
            (2\kappa r) + \gamma _{j} +\frac{\sigma _{1} - \sigma _{2}}{2}
            \right]-\nonumber \\&& \frac{|\lambda _{j} + i \beta|}{2 \kappa
            r} \cos \left[\kappa r -\lambda _{j} \frac{\pi}{2} - \beta ln(2
            \kappa r) + \gamma _{j} +\frac{\sigma _{1} + \sigma _{2}}{2}
            \right]+...\Bigg\}
\end{eqnarray}
where
\begin{eqnarray}
\label{63}
\sigma_{1} = arg\;\; \left( \frac{1}{\lambda_{j} - i \beta}\right) ,
\end{eqnarray}
\begin{eqnarray}
\label{64}
\sigma_{2}= arg\;\;  \left[ \frac{1}{-\eta/ \alpha - iMK/ \kappa \alpha^
        {2} }\right] 
\end{eqnarray}
and
\begin{eqnarray}
\label{65}
\gamma_{j} = arg\Gamma(\lambda_{j} + i \beta).
\end{eqnarray}

From the above expressions we can obtain the phase shift, which is given by
\begin{eqnarray}
\label{66}
\delta_{j} = \frac{\pi}{2}[(j + 1/2) - \lambda_{j}] + \gamma_{j} 
           + \frac{\sigma_{1} - \sigma_{2}}{2}.
\end{eqnarray}

As in bosonic case we can see that the phase shift presents contibuitions
comming from :
$(i) $ The modification in the effective total angular quantum numbers, 
$\lambda_{j} $, due to the geometry of the manifold it self and $ (ii) $
the presence of the induced self-interaction. There is also an additional
contribution when we compare $ (\ref{63}) $ with $ (\ref{33}), $  the
phase shift for the bosonic scattering states, this difference is due to
specific behaviour of the spinor field in presence of an external
potential.
\newpage
\renewcommand{\thesection}{\arabic{section}.}
\section{\bf Concluding Remarks}
$   $
\,\,\,\,\,\,\,In this paper we have analysed the relativistic motion at 
classical and quantum point of view, of a charged particle in the neighbourhood
of a global monopole, considering the induced electrostatic  self-
interaction as the zeroth component of an external four vector potential
$ A_{\mu} $. Although the magnitude of the self-interaction can be a small
quantity for a typical model of global monopole system comming from a grand
unifield theory \footnote{For a typical grand unifield theory the
parameter $ \eta, $ associated with the scale where the global symmetry is
spontaneously broken, is of the order $ 10^{16} $ Gev and $ \triangle =1 -
\alpha^{2} = 8 \pi G \eta^{2} \simeq 10^{-5} $ . In Ref.$[2]$  it is
estimated the magnitude of the induced self-interaction for small $
\triangle $, it is approximately $ K\approx \frac{q^{2}\pi \triangle}{32}.
$}, which seems to be the more problabe, its effects can be measured by the
phase shift for scattering elementary particle by a global monopole.

In order to make our analysis as more complete as possible, we decided to
investigate the behaviour of the classical and quantum motion  of the 
charged particle, considering the possibility to have an attractive
electrostatic self-interaction.For this case we could observe that it
is possible to construct bound states for the system composed by the 
particle and the global monopole.

The main results of this paper were the obtainment of exact equation for
classical orbits, self-energies, phase shifts and complete eigen-function
at quantum level, for the movement of a charged particle in the
neighbourhood of a global monopole.

Finally we would like to make a comment about the procedure adoptel by us
in this paper. Although we have used the induced electrostatic self-
interaction given in Eq.$ (\ref{3}) $ to study the relativistic motion a
charged particle, this result was obtained in Ref.$ [2] $ assuming that the 
particle was at a fixed position outside the global monopole. The complete
procedure to determine the self-interaction due to a moving particle in the
presence of a global monopole,  should take into account the previous
knowledgement of the particle's trajectory. Because this procedure makes
a concret analysis about this system almost impossible we decided to adopt
the quasistatic approach presented here, i.e, we neglected the corrections
on the self-interaction comming from this movement  \footnote{ In Ref.$[10]
$ is presented the induced electromagnetic semmmlf-interaction on a charged
particle moving along an arbitrary trajectory on a conical space-time.}.
\\      \\
{\bf{Acknowledgments}}
\\       \\

We would like to thank Conselho Nacional de Desenvolvimento Cient\'{\i}fico
e Tecnol\'{o}gico (CNPQ) and CAPES for partial financial support. A. A. 
Rodrigues Sobreira also thanks the Universidade Estadual da Para\' {\i}ba.


\newpage
\begin{thebibliography}{100}
\bibitem{barriola} Barriola M and Vilenkin A, 1989 {\it Phys.Rev.Let}.{\bf
63}, 341.
\bibitem{eugenio} Bezerra de Mello E R  and  Furtado C, 1997  {\it Phys.
Rev.}{\bf D56},1345.
\bibitem{landau} Landau L  et Lifchitz E, 1966 {\it Th\'{e}orie du Champ}, 
\'{E}ditions Mir(Moscou,). Ch 5
\bibitem{chakraborty} Chakraborty S, 1996 {\it General Relativistic and 
Gravitation} {\bf 28},9.
\bibitem{gross} Gross F, 1993 {\it Relativistic Quantum Mechanics and Field 
Theory}, ed. John Willey.
\bibitem{schiff} Schiff L. I. 1968 {\it Quantum Mechanics}, 3rd ed(Mc.Graw
-Hill International).
\bibitem{mazur} Mazur P. O. and Papavassiliou J.1991, {\it Phys.Rev.}{\bf
D44}, 1317.
\bibitem{bjorken} A similar set of equations is obtained for the analysis
of the relativistic hidrogen atom in J. D. Bjorken and S. D. Drell, {\it
Rellativistic Quantum Fields}, McGraw-Hill Book Co, 1965.
\bibitem{abramowitz} Abramowitz M  and Stegun I A 1970, {\it Handbook of 
Mathematical Functions}(New York:Dover).
\bibitem{nk} Khusnutdinov N R 1994 {\it Class Quantum Gravitational} {\bf
11},1807.
\end{thebibliography}
\end{document}